\documentstyle[11pt,epsfig]{article} 
\title{Southern Hemisphere Observations of a $10^{18}$eV Cosmic Ray Source Near the Direction of the Galactic Centre}
\author{J.A.\,Bellido, R.W.\,Clay, B.R.\,Dawson and M.\,Johnston-Hollitt\\Department of Physics\\The University of Adelaide, Australia 5005\\}

\begin{document}
\begin{titlepage}
\maketitle
\begin{abstract}

We report on an analysis of data from the southern hemisphere
SUGAR cosmic ray detector.  We confirm the existence of an excess
of $10^{18}$eV cosmic rays from a direction close to the
Galactic Centre, first reported by the AGASA group.  We find that
the signal is consistent with that from a point source, and we
find no evidence for an excess of cosmic rays coming from the
direction of the Galactic Centre itself.

\end{abstract}
\centerline{{\em Submitted to Astroparticle Physics, June 5 2000}}
\centerline{{\em Revised July 11 2000}}
\centerline{{\em Accepted July 29 2000}}

\end{titlepage}

\section{Introduction}

It is generally accepted that the majority of cosmic rays
observed on Earth are accelerated within our galaxy.  However,
apart from our Sun at low energies, no specific acceleration site
has previously been confidently identified for hadronic cosmic
rays.  Various mechanisms have been proposed for their
acceleration, including the very efficient shock acceleration in
supernova explosions.  All proposed galactic mechanisms have
difficulty explaining the highest energy cosmic rays, and the
lack of any likely candidates has led to a search for
extragalactic sources for these particles.  The energy at which
galactic sources might be overtaken by extragalactic sources is
not at all clear but is of considerable interest.  The situation
is complicated by the galactic magnetic field which is thought to
be strong enough (of order $1\mu$G) to scramble the directions of
charged cosmic rays at energies up to at least $10^{18}$eV
\cite{Roger}.  It is only at $10^{19}$eV, when the Larmor radius
of a cosmic ray proton becomes comparable with the galactic scale,
that one can confidently expect a clear galactic anisotropy if
galactic protons exist at these energies.  No strong cosmic ray
anisotropy had been detected from galactic directions until very
recently.

In 1999 the AGASA group reported a study of data collected with
their 20\,km$^2$ and 100\,km$^2$ extensive air shower arrays over
the period from 1984 to mid-1995 \cite{AGASA99}.  With such a
large data set they were able to systematically perform harmonic
analyses in right ascension as a function of energy, from
$10^{17}$eV to $10^{20}$eV.  They discovered a large first
harmonic anisotropy (amplitude 4\%) over the narrow energy range
from $10^{17.9}$ to $10^{18.3}$eV.  A sky map of the cosmic ray
density showed that this anisotropy was apparently caused by an
excess in a direction close to the Galactic Centre, with a
possible smaller excess from the direction of the Cygnus region
of the galactic plane.  When producing the sky map, the AGASA
group tried 4 different assumptions for the signal beam size, and
found that the near-Galactic Centre excess was most significant
when a beam size of radius $20^\circ$ was assumed.  That excess
was a $4.1\sigma$ deviation above the expected isotropic flux.

This very important analysis was complicated by the number of
trials in energy and beam size used to produce the final result.
More recently the group has completed an analysis of AGASA data
up to April 1999 \cite{ICRC99}.  This extra time has resulted in
almost a doubling in the number of events in the energy region
around $10^{18}$eV.  A new analysis was performed on the entire data
set from 1984 to 1999, and the results of the original study were
confirmed.  The group selected a slightly different energy range,
from $10^{18.0}$ to $10^{18.4}$eV, and produced a sky map assuming a
beam size of radius $20^\circ$.  They found a $4.5\sigma$ excess
near the Galactic Centre (506 events with a background of 413.6
events), with a smaller $3.9\sigma$ excess (3401 events with a
background of 3148) seen in the Cygnus region.

Some evidence for a galactic plane excess was also seen in data
from the Fly's Eye experiment over a similar energy range
\cite{Alfred}.  This was a broad-scale study, and no attempt was
made to identify whether any particular galactic longitudes were
responsible for the excess.  As pointed out by the AGASA group,
other experiments such as Haverah Park and Yakutsk are too
northerly to see the excess region identified by AGASA near the
Galactic Centre.

The energy dependence of the anisotropy may offer a clue to the
nature of the signal particles.  At an energy of $10^{18}$eV a
neutron will have a decay length of approximately 10\,kpc,
roughly the distance to the Galactic Centre \cite{Jones}.  At
energies lower than this, neutrons from the distance of the
Galactic Centre would decay before reaching Earth.  The
appearance of the anisotropy at around $10^{18}$eV could be
explained by this effect.  Its disappearance at $3\times10^{18}$
might then be explained by the source reaching its energy limit.
Other clues to the nature of the signal particles would include
the scale of the anisotropy on the sky (point source-like or
broad) and the response of different detector arrays to the
signal events (a possible discriminant between hadronic and
non-hadronic primary cosmic rays).

\section{The SUGAR Air Shower Detector}

Because of the detector latitude and a shower zenith angle cut
($60^\circ$), the AGASA sky map cuts off at a declination of
$-24.2^\circ$, some $4.7^\circ$ north of the Galactic Centre
$(\alpha,\delta)=(266.4^\circ,-28.9^\circ$) (J2000.0).  Their most
significant excess, near the Galactic Centre, is on the edge of
the sky map.  In order to study this region and more southerly
declinations, we have used data collected by a large southern air
shower array, using the AGASA results to define {\em a priori}
cuts and thus avoiding statistical penalties.

The SUGAR array was located in the Pilliga State Forest in New
South Wales, Australia (array latitude 30.53$^\circ$S , longitude
149.60$^\circ$E) and was operated between 1968 and 1979. It
enclosed an area of up to 70 km$^{2}$, and was designed to detect
showers produced by the highest energy cosmic rays \cite{Winn86}.
The array had an energy threshold of about $2\times10^{17}$eV,
and in its final configuration consisted of 47 independent
stations. Each station contained two buried liquid scintillator
counters separated by a distance of 50\,m. The shielding over
each detector was $(1.5\pm0.3)$m of earth, and hence these
detectors were sensitive to the penetrating muon component of
extensive air showers (with a muon energy threshold 0.75\,GeV for
vertical showers).

The typical detector station spacing in the array was 1.6\,km,
since the main purpose of the array was to study very large
showers with energies above $10^{19}$eV.  However, a small
(approximately 1\,km$^2$) sub-array with detector spacing of 500\,m was
operated with sensitivity to showers around $10^{18}$eV.

\section{Analysis}

Our intention was to study the Galactic Centre region highlighted
by the AGASA analysis, and we have used their results to set {\em
a priori} the energy range to be used in this SUGAR analysis.
The SUGAR group determined the primary particle energy from a
shower's equivalent vertical muon size by applying a conversion
factor determined by early simulations of Hillas (see
\cite{Winn86}).  We have found that the SUGAR integral energy
spectrum derived with this technique is in good agreement with
the AGASA spectrum \cite{AGspect} around $10^{18}$eV.  We have
chosen ({\it a priori}) an energy range of $10^{17.9}$ to
$10^{18.5}$eV for our study.  This is slightly wider than the
optimal AGASA range because we recognise that the SUGAR energy
resolution is likely to be poorer than that of AGASA.

The SUGAR data set yielded 3732 events within this energy range.
We produced a shower density sky map using the following
technique.  Each event was assumed to have a directional
uncertainty of $3^\circ \sec\theta$, where $\theta$ is the zenith
angle of the event.  This uncertainty is based on a study of
events triggering more than three SUGAR stations \cite{Winn86}.
We have applied it to all events in our sample (54\% of which
triggered more than three detector stations) as the best
available estimate of the uncertainty.

We represented each event by a gaussian probability function
surrounding its nominal direction on the sky, with the gaussian's
standard deviation equal to the directional uncertainty.  The
volume of the gaussian was normalized to one.  All events were
added to the map in this way, resulting in a shower density map
(Figure 1).

We next compare this map with the expectation based on an
isotropic flux of cosmic rays.  That expectation must take into
account the exposure of the array in right ascension and
declination.  It was determined using the ``shuffling'' technique
\cite{Sommers}.  Here, a number of shuffled data sets are derived
from the real data set, with each shuffled data set containing
the same number of events as the real one.  A real arrival time
(Julian date) from one event is randomly paired with a local
arrival direction (zenith angle, azimuth angle) from another
event in the real data set.  This is repeated until a new data
set is filled.  The new data set has the same arrival time
distribution and the same zenith and azimuth angle distributions
as the real data set.  However, because the pairings have been
randomized, all celestial directions have been randomized
producing an isotropic event flux.  Many shuffled data sets can
be generated.  For each of those, an event density map can be
generated (like Figure 1) using gaussian point-spread functions.
Figure 2 shows the isotropic background shower density map,
derived from an average of 1000 shuffled maps.

Comparing the real density map with the isotropic expectation, we
derive a map showing the fractional excess of event densities
across the sky, shown in Figure 3.  We show this map to emphasise
that regions of excess are rare.  Only two regions of interesting
excess are apparent, with one of them near-coincident with the
strongest AGASA excess.  The other excess is quite likely to be a
statistical fluctuation, given the number of unique directions
represented on the map.  This second region is not of interest to
us as it was not part of our analysis strategy derived {\em a
priori} which related to the region of the AGASA excess.  The
interesting SUGAR excess is centered at
$(\alpha,\delta)=(274^\circ,-22^\circ$) (B1950.0), close to the
position of the AGASA excess (see below).

We next calculate the statistical significance of the SUGAR
excess.  To do this we again use the shuffled data sets.  We grid
the original shower density map (Figure 1) into $0.5^\circ$ bins,
and ask how many of the 1000 shuffled data sets have a shower
density in the bin equal to or larger than the real bin density.
Given that each shuffled map is a representation of an isotropic
cosmic ray flux, this gives us a bin-by-bin probability that the
real map density has occurred by chance.

We show the significance map for the signal region in Figure
4.  The contours represent chance probabilities, with the signal
region peaking at a chance probability of 0.005.  (On this map,
regions of excess will have chance probabilities less than 0.5, and
regions of deficit will have chance probabilities greater than 0.5).

\section{Discussion}

Figure 4 shows at least two interesting features.  First, there
is no hint of a signal from the true centre of the galaxy, even
though SUGAR (unlike AGASA) had a clear view of this region.  The
peak of the signal region is $7.5^\circ$ from the Galactic
Centre.  Secondly, the signal region is no larger than would be
expected for a point source of cosmic rays.  The size of the
region is consistent with the detector's angular precision.  This
would suggest that the particle source is not an extended region,
and that the particles have not experienced large direction
deviations during propagation.

In Figure 5 we compare our result with the AGASA map, using the
2,3 and $4\sigma$ contours from reference \cite{ICRC99}.  The
peaks of the signals are displaced by about $6^\circ$, and the
size of the signal region is clearly different with AGASA seeing
a much broader enhancement.  There are several possible reasons
for the differences, including

\begin{itemize}
\item The SUGAR data is, simply by chance, tightly clustered and
offset from the AGASA signal.
\item The SUGAR peak could be offset due to a systematic pointing
error in local coordinates.  This question has been thoroughly
investigated by the SUGAR group \cite{Brownlee}.  Possible
sources of systematic errors in pointing were investigated (array
survey, curvature of the Earth, electronics, propagation of radio
timing signals through a refractive atmosphere and vegetation,
the effect of ghosting of these signals etc) and it was concluded
that likely systematic errors were small compared with 
random errors.  In any case, a systematic pointing error of
reasonable magnitude would also cause a significant smearing of
the signal in right ascension and declination.
\item The AGASA angular resolution is poorer than expected at
large zenith angles, and/or AGASA suffers from a systematic pointing
error which manifests itself as a smearing and offset of the
signal in right ascension and declination.
\item The AGASA analysis technique, which uses signal averaging
over $20^\circ$ radius circles, has smeared out an otherwise
point-like feature.
\item Apparent excesses on the AGASA map further north along the
galactic plane might have systematically shifted the peak of
their main excess northward.  No similar bias would have an
opportunity to act southward, given their cutoff in declination
at $\delta=-24^\circ$ .
\end{itemize}
We encourage the AGASA group to consider performing an analysis
of their data with a technique similar to ours.  This technique
requires an estimate of arrival direction uncertainty for each
event, but it removes the need to search for an optimal beam size
and it uses the data themselves to estimate the background.

We calculate the SUGAR source flux in the following way.  Given our
assumption that the size of the signal region is consistent with
point-spread function, we count the number of events inside a
circle of radius $5.5^\circ$ centered on
$(\alpha,\delta)=(274^\circ,-22^\circ$).  (The average zenith
angle for events at this declination is approximately $30^\circ$,
so the typical direction uncertainty would be
$3^\circ\sec(30^\circ)$.  We multiply this radius by 1.59 in
order to maximize the signal-to-noise ratio of our flux
estimate).  We find that this region of the density map contains
21.8 equivalent events.  This is compared with the background
expectation of 11.8, the average number of events inside 27
circles of the same size arranged around same declination band.
Thus the signal region is populated by approximately 10 signal
events above an expectation of 11.8 background events.  This is
consistent with the chance probability of 0.005 derived above
using shuffling and the sky maps. (The Poisson probability of
observing 22 or more events when the expectation is 11.8 is
0.0050).  Using the AGASA energy spectrum \cite{AGspect} as
normalization, and assuming that the signal events have the same
triggering efficiency as the background cosmic rays, this excess
is equivalent to a point-source flux of $(9\pm3)\times10^{-14}$
m$^{-2}$s$^{-1}$ or $(2.7\pm0.9)$ km$^{-2}$yr$^{-1}$ between
$10^{17.9}$ and $10^{18.5}$eV.  We cannot compare this flux with
the AGASA result \cite{ICRC99}, since no flux is quoted and we do
not know the procedure used by the AGASA group to estimate the
signal and background counts within the large ($20^\circ$ radius)
error circle used in their analysis.  In particular, such a
circle centred on the excess extends beyond the southern limit of
the AGASA map.

The excess shown in Figure 4 is clearly not coincident with the
Galactic Centre.  At the declination of the Galactic Centre, the
expected number of events between $10^{17.9}$ and $10^{18.5}$eV
arriving within a circle of radius $5.5^\circ$ is 13.4.  The
actual number of events within $5.5^\circ$ of the Galactic Centre
is 12.5.  Thus, we calculate a 95\% upper limit \cite{Protheroe}
on the point source flux from the Galactic Centre of 2.2
km$^{-2}$yr$^{-1}$ (or roughly 70\% of the cosmic ray flux) over
this energy range.

The peak of the SUGAR excess does not appear to be centered on
the galactic plane itself.  However, at a galactic latitude of
only about $3^\circ$ south, it is still within the SUGAR angular
uncertainty of the plane.

We have examined astronomical data in the direction of the SUGAR
peak to determine whether there might be any coincident objects
of interest.  The direction is on the border of galactic
plane surveys and many surveys have only statistically poor
information.  However, an 11cm Effelsberg 100m single dish radio
survey (sensitive to broad scale structure) \cite{Effelsberg} and
data from the COMPTEL gamma-ray telescope \cite{Schon96} do seem
to be useful, based on their angular coverage and the apparent
quality of the data.

The 10-30 MeV COMPTEL dataset shows a large feature which arcs
south of the galactic plane, from the Galactic Centre to an
unidentified bright source in the galactic plane at $18^{\circ}$
galactic longitude.  That arc, with a radius of about seven
degrees, is apparently centred on the direction of the SUGAR
peak.  The radio data show a bright region with a radius of about
one degree in the direction of the peak and this appears to be
surrounded by a roughly circular feature of much reduced radio
intensity, again centred on the peak.  There may be a brighter
radio region outside that feature and coincident with the COMPTEL
arc.  The central radio source is polarised but there is no sign
in the data for radio polarisation associated with the COMPTEL
arc. If this feature was at the distance of the Galactic Centre,
it would have a diameter of about 800pc.

We emphasise that these other astronomical data are within the
SUGAR angular resolution and their apparently coincident
directions may be a statistical artefact.

\section{Conclusion}

Data from the SUGAR array confirm the existence of an excess flux
of cosmic rays from a direction near the Galactic Centre.  While
this result is not as statistically strong as that reported by
the AGASA group, it is interesting in a number of ways.  First,
the SUGAR array consisted of buried scintillator detectors, with
a muon energy threshold of 0.75\,GeV for vertical showers.  If
the SUGAR flux we calculated above proves to be consistent with
that measured by AGASA, it would imply that the signal particles
are unlikely to be gamma-rays, unless our understanding of muon
production in photon cascades is severely incomplete.  

Secondly, the SUGAR array had a near overhead view of the true
Galactic Centre, and found no signal from that direction.  This,
coupled with the observation that the SUGAR signal is point-like
in character, raises the possibility that the source of these
particles is unrelated to the centre of our galaxy.  For example,
it is conventional to think of the galactic magnetic field as a
superposition of regular and turbulent components.  It would be
difficult to conceive of a field structure which would take a
source of charged cosmic rays at the Galactic Centre and make it
appear like a point source offset by $7.5^\circ$ from the true
source direction.  Clay \cite{Roger2000} has discussed
propagation from such a source and has shown that a large diffuse
region would result, a region much larger than the point spread
dimensions observed with SUGAR.

The possibility of neutron primary particles cannot be ignored,
especially as this could account for the turn-on of the signal at
around $10^{18}$eV (e.g. \cite{Jones}) if the source were at a
distance close to that of the galactic centre.  It would also
account for the point-like character of the excess as seen by
SUGAR.

\section{Acknowledgements}

We acknowledge the pioneering work carried out by our colleagues
at the University of Sydney who built and operated the SUGAR
array, and thank them for access to their data.  We are grateful
for very useful discussions with M.M. Winn and J. Ulrichs from
Sydney, and G.J. Thornton from our own institution.  This work is
supported by the Australian Research Council.

\begin{figure}[p]
\begin{centering}
\epsfig{file=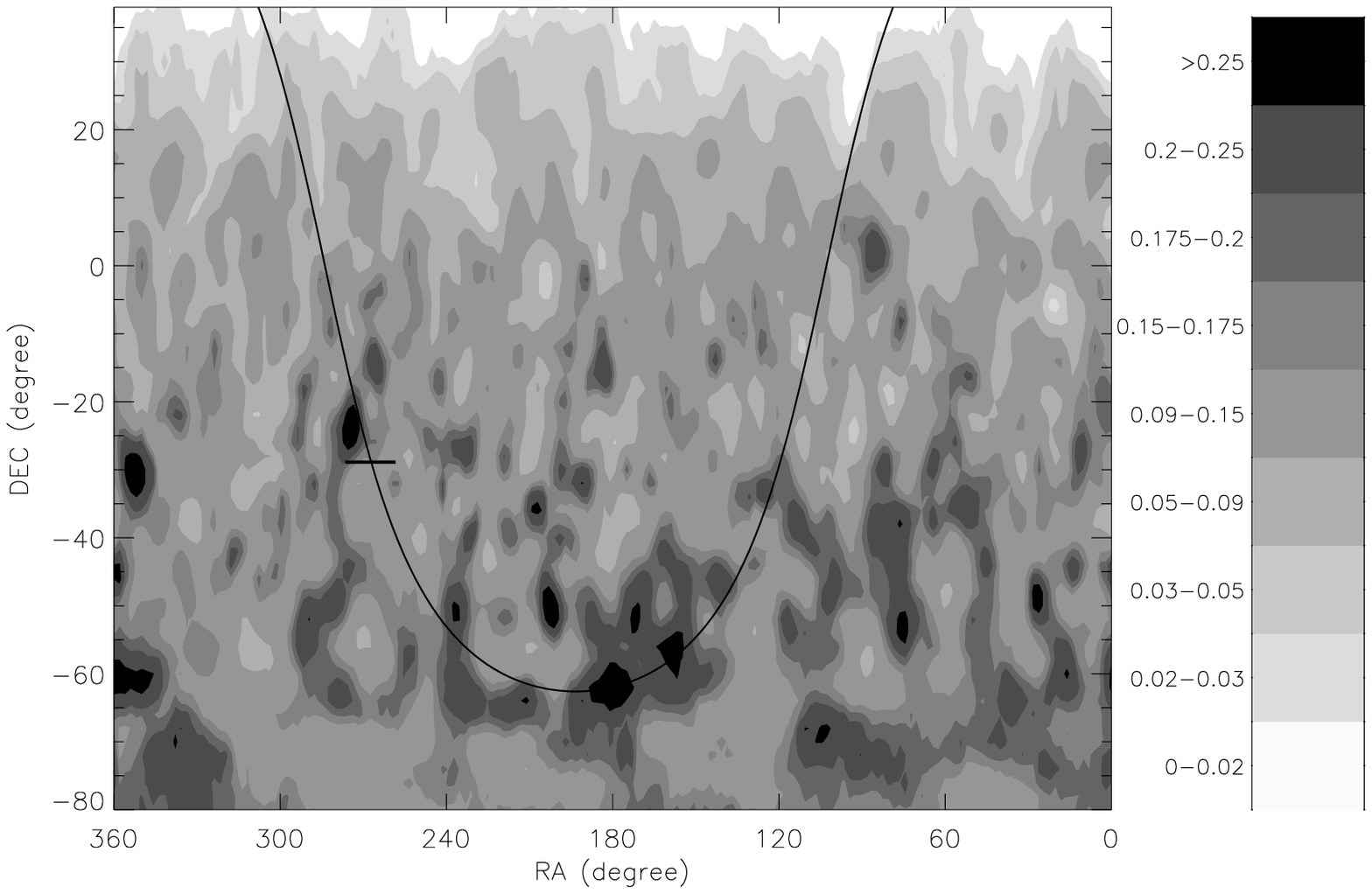,width=16cm,angle=0}
\caption{Cosmic ray event density over the sky viewed by the
SUGAR array for the 3732 events between $10^{17.9}$eV and
$10^{18.5}$eV.  The density scale represents the number of events
viewed per true square degree of sky, where a true square degree
is defined as being 1 degree in declination $(\delta)$ $\times$
$1^\circ\sec\delta$ in right ascension.  A point spread function
has been applied to every event to represent the angular
uncertainty in their arrival directions.  The galactic plane and
Galactic Centre are indicated with the solid line and cross
respectively.  The 1950 epoch has been assumed for the equatorial
coordinates displayed here and in other plots.}
\end{centering}
\label{fig:1} 
\end{figure} 

\begin{figure}[b]
\begin{centering}
\epsfig{file=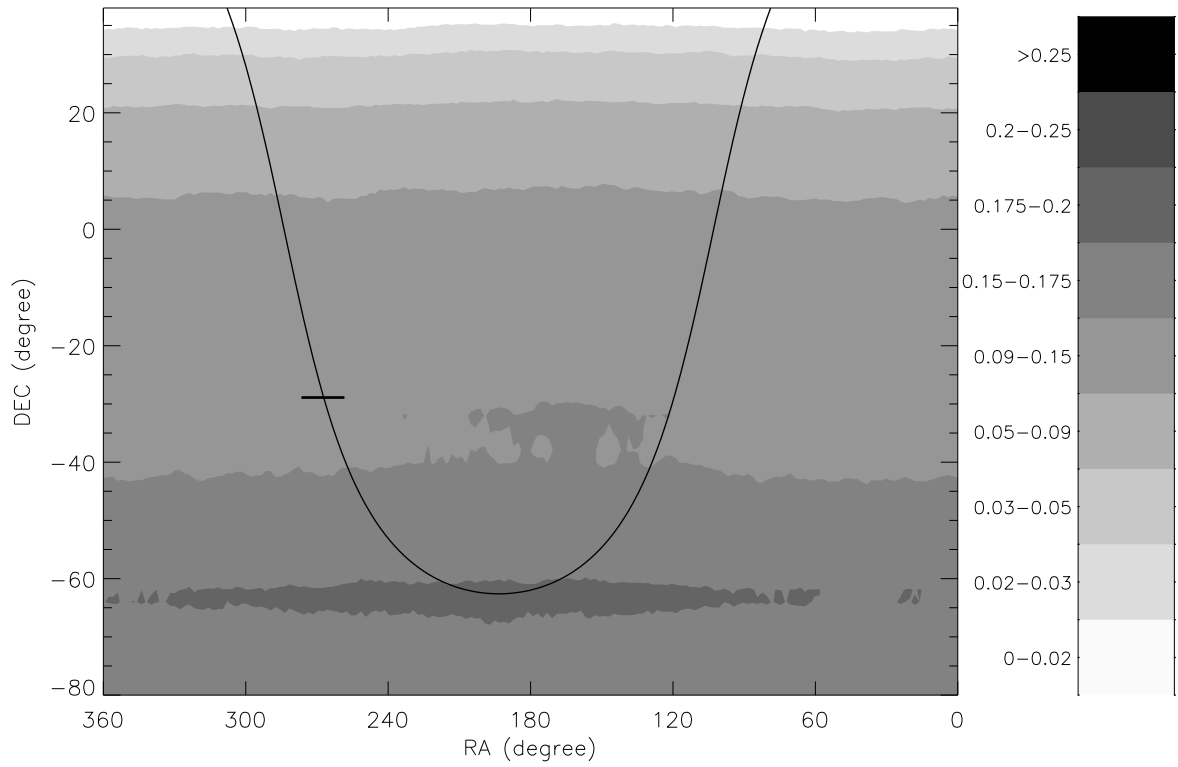,width=16cm,angle=0}
\caption{The expected density of events for an isotropic flux of
cosmic rays as viewed by SUGAR between $10^{17.9}$eV to
$10^{18.5}$eV.  Again, the density is given in units of events
per true square degree (see caption to Figure 1).}
\end{centering}
\label{fig:2} 
\end{figure} 

\begin{figure}[t]
\begin{centering}
\epsfig{file=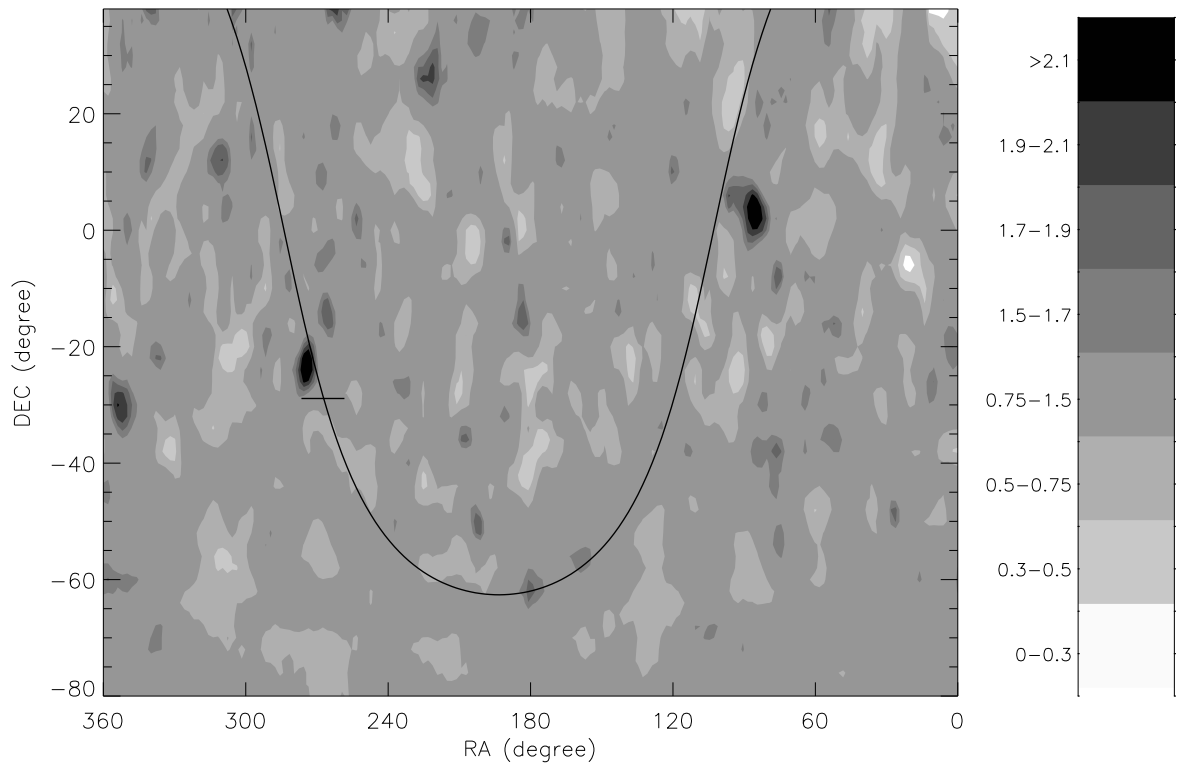,width=16cm,angle=0}
\caption{We compare Figures 1 and 2 and derive the fractional
excess (or deficit) of the event density over the sky viewed by
SUGAR.  A value of 1 indicates that the measured density is in
agreement with the expected density.}
\end{centering}
\label{fig:3} 
\end{figure} 

\begin{figure}[t]
\begin{centering}
\epsfig{file=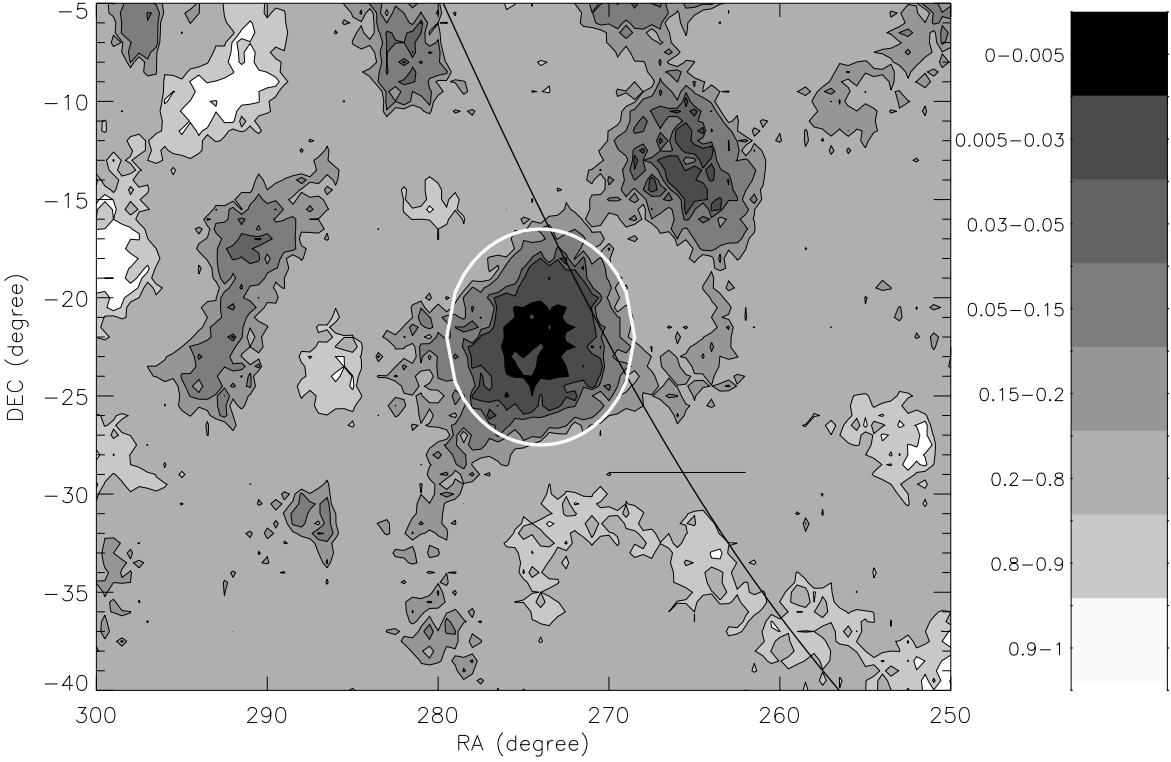,width=16cm,angle=0}
\caption{The significance of the excess detected in the Galactic
Centre region, as calculated using shuffled data sets.  The
contours represent the chance probability of SUGAR detecting the
observed density or greater.  Thus, a contour level of 0.5
represents a measured density which is consistent with
expectation.  The peak of the signal region has a chance
probability of 0.005.  The galactic plane and Galactic Centre are
indicated by the solid line and cross respectively.  The white
circle of radius $5.5^\circ$ represents the typical error circle
for SUGAR events.  The excess therefore appears to be consistent
with that from a point source.}
\end{centering}
\label{fig:4} 
\end{figure} 

\begin{figure}[t]
\begin{centering}
\epsfig{file=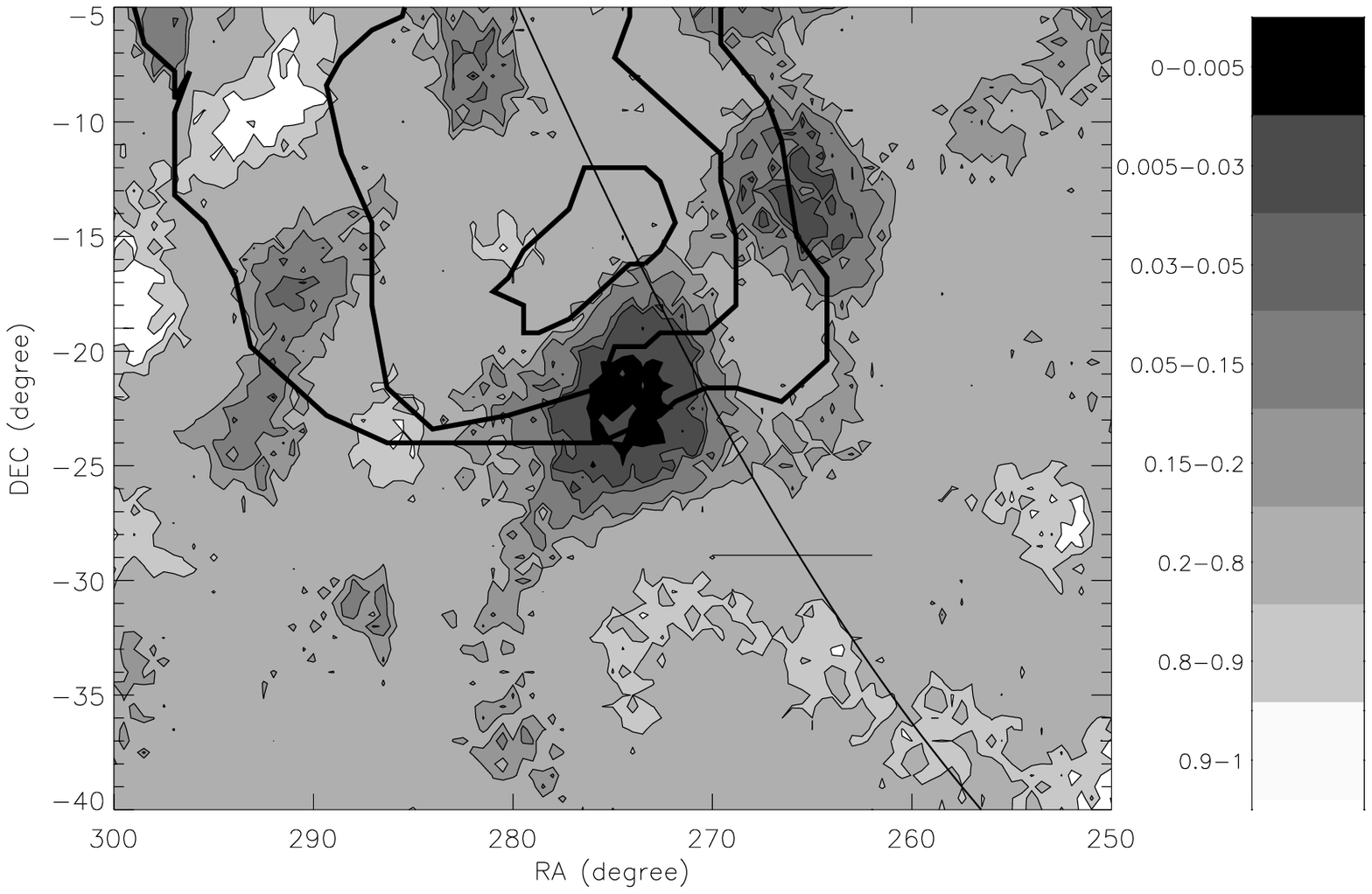,width=16cm,angle=0}
\caption{A comparison of the AGASA and SUGAR results.  The SUGAR
map from Figure 4 is overlaid with $2\sigma$, $3\sigma$ and
$4\sigma$ contours from reference \cite{ICRC99}.  Note that the
limit of AGASA's view is close to $\delta=-24^\circ$, indicated by the
horizontal portion of the $2\sigma$ contour.  The AGASA
signal region size is significantly larger than seen by SUGAR.
See the text for a discussion.}
\end{centering}
\label{fig:5} 
\end{figure}


\begin{thebibliography}{999}
\bibitem{Roger} A.A. Lee and R.W. Clay, J.Phys.G:Nucl.Phys.,
{\bf{21}} (1995) 1743.
\bibitem{AGASA99} N. Hayashida et al., Astropart. Phys. {\bf 10} (1999) 303.
\bibitem{ICRC99} N. Hayashida et al., Proceedings of the $26^{\rm
th}$ Int. Cosmic Ray Conf., Salt Lake City {\bf 3} (1999) 256.
\bibitem{Alfred} D.J. Bird et al., Ap.J. {\bf 511} (1999) 739.
\bibitem{Jones} L.W. Jones, Proc. 21st Int. Conf. Cosmic Rays (Adelaide),
University of Adelaide, {\bf 2} (1990) 75
\bibitem{Winn86} M.M. Winn et al., J.Phys.G:Nucl.Phys., {\bf{12}} (1986) 653.
\bibitem{AGspect} M. Nagano et al., J.Phys.G:Nucl.Phys., {\bf{18}} (1992) 423.
\bibitem{Sommers} G.L. Cassiday et al., Nucl. Phys. B (Proc. Suppl.) {\bf{14A}}
(1990) 291.
\bibitem{Brownlee} R.G. Brownlee, ``A Directional Analysis of Very High
Energy Air Showers'', Ph.D. Thesis, University of Sydney (1970).
\bibitem{Protheroe} R.J. Protheroe, Astronomy Express {\bf{1}} (1984) 33.
\bibitem{Effelsberg} W. Reich et al., in ``New Perspectives on
the Interstellar Medium'', ASP Conference Series 168 (A.R. Taylor
et al. eds) Astron. Soc. of the Pacific, 78 (1999).
\bibitem{Schon96} V. Sch\"{o}nfelder et al., Astron. Astrophys. Suppl. Ser {\bf 120} (1996) 13.
\bibitem{Roger2000} R.W. Clay, ``The Propagation of Cosmic Rays
from the Vicinity of the Galactic Centre'',
Publ. Astron. Soc. Aust. (in press) (2000).

\end{thebibliography}
\end{document}